\documentclass[10pt, conference, letterpaper]{IEEEtran}
\usepackage{cite}
\usepackage{graphicx}
\usepackage{balance}
\usepackage{epsfig}
\usepackage{epstopdf}
\usepackage{enumitem}
\usepackage{wrapfig}
\usepackage{amsmath}
\usepackage{tabularx}
\usepackage{makecell}
\usepackage{algorithm}
\usepackage{algorithmic}
\usepackage{amssymb}
\usepackage{amsfonts}

\ifCLASSINFOpdf
\else 
\fi
\usepackage{colortbl}

\usepackage{lipsum}
\ifCLASSOPTIONcompsoc
    \usepackage[caption=false, font=normalsize, labelfont=sf, textfont=sf]{subfig}
\else
\usepackage{caption}
\usepackage{subcaption}
\fi
\newcommand{\pol}[0]{\pmb{\pi}}
\newcommand{\cpol}[0]{\pmb{\mu}}
\newcommand{\Cpol}[0]{\pmb{\mathcal{M}}}

\newcommand{\classifier}[0]{\mathcal{C}}

\begin{document}
\title{{Cascade Reinforcement Learning with State Space Factorization for O-RAN-based Traffic Steering}}
\author{{Chuanneng Sun, Gueyoung Jung, Tuyen X. Tran, Dario Pompili}\\
Dept. of Electrical and Computer Engineering, Rutgers University--New Brunswick, NJ, USA\\
\textit{\{chuanneng.sun, tuyen.tran, pompili\}@rutgers.edu; gueyoung.jung@gmail.com}
}

\maketitle
\pagenumbering{arabic}

\begin{abstract}
We study the Traffic Steering~(TS) problem in Open Radio Access Network~(O-RAN), leveraging its RAN Intelligent Controller~(RIC), in which RAN configuration parameters of cells can be jointly and dynamically optimized in near-real-time.
To address the TS problem, we propose a novel Cascade Reinforcement Learning~(CaRL) framework, where we propose state space factorization and policy decomposition to mitigate the need for large complex models and well-labeled datasets. For each sub-state space, an RL sub-policy is trained to optimize the Quality of Service~(QoS). To apply CaRL to new network areas, we propose a knowledge transfer approach to initialize a new sub-policy based on knowledge learned by the trained policies. 
To evaluate CaRL, we build a data-driven and scalable RIC Digital Twin (DT) that is modeled using real-world data, including network setup, user geo-distribution, and traffic demand, among others, from a tier-1 RAN operator. We evaluated CaRL in two DT scenarios representing two different US cities and compared its performance with business-as-usual policy as a baseline and other competing optimization approaches (i.e., heuristic and Q-table algorithms). Furthermore, we have conducted a field trial with the RAN operator to evaluate the performance of CaRL in two areas in the Northeast US regions.
\end{abstract}

\IEEEpeerreviewmaketitle

\vspace{-.1in}
\section{Introduction} \label{sect:intro}

\begin{figure*}[t!]
    \centering 
    \begin{subfigure}{0.22\textwidth}
        \includegraphics[width=\linewidth]{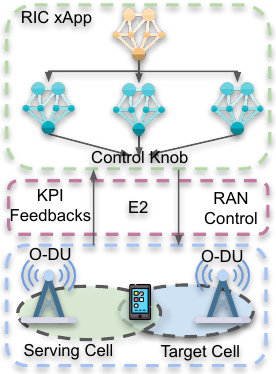}
        \caption{\label{fig:problem_demo}}
    \end{subfigure}\hfil 
    \begin{subfigure}{0.3\textwidth}
        \includegraphics[width=\linewidth]{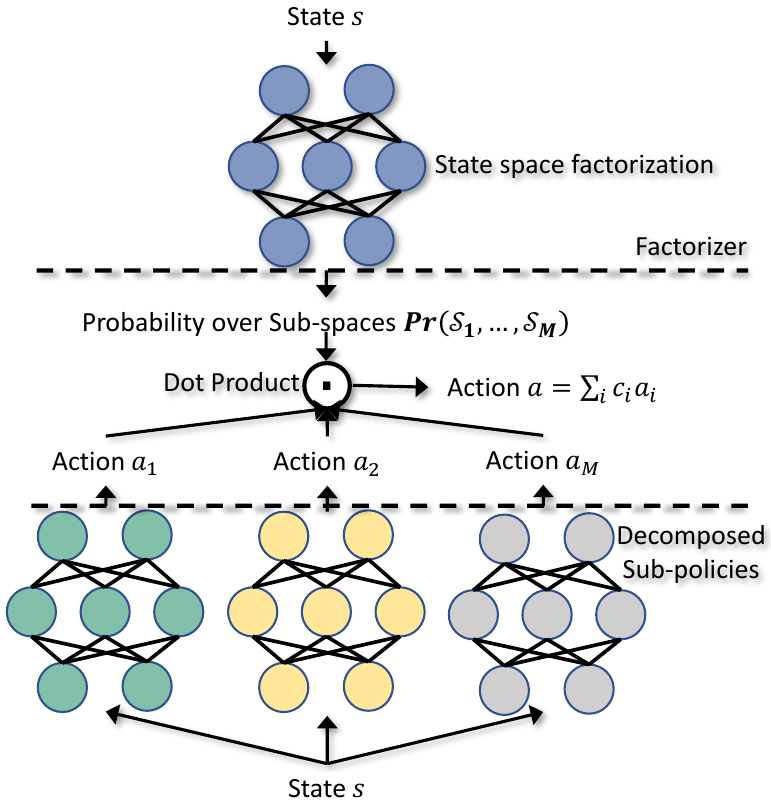}
        \caption{\label{fig:CaRL}}
    \end{subfigure}\hfil 
    \begin{subfigure}{0.35\textwidth}
        \includegraphics[width=\linewidth]{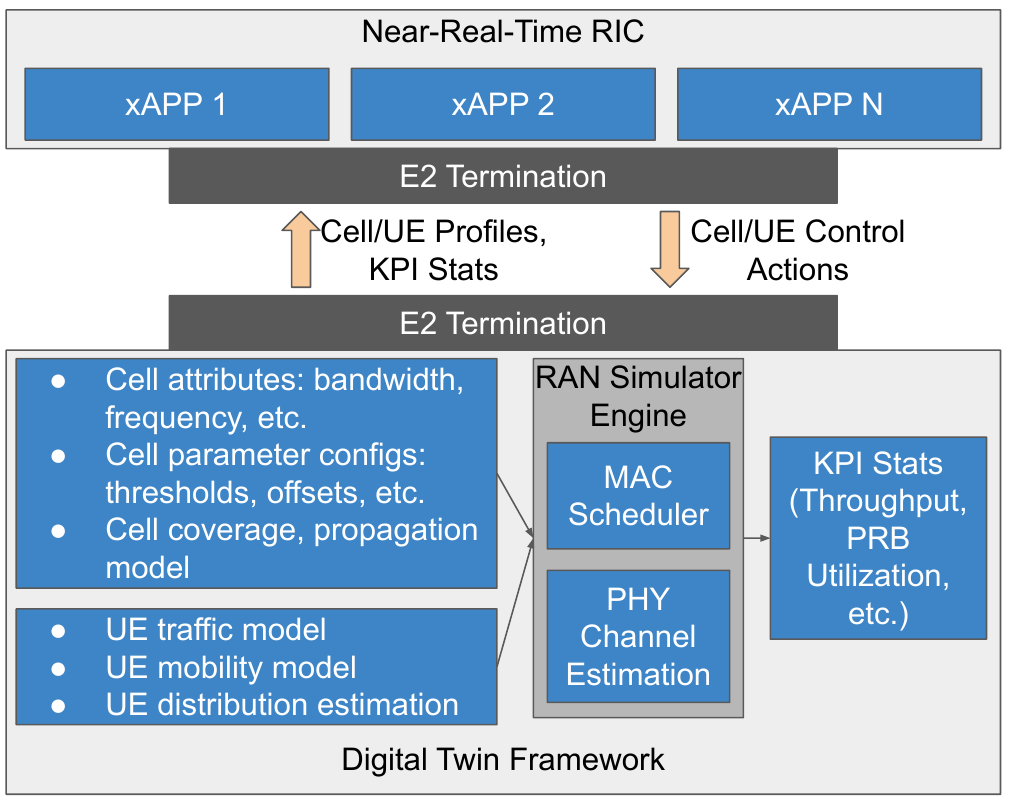}
        \caption{\label{fig:sim}}
    \end{subfigure}\hfil 
    \caption{\label{fig:combined} (a) A visualization of the O-RAN facilitated traffic steering (TS) application. (b) Structure for the cascade policy. There are $M$ sub-policy neural networks corresponding to $M$ sub-spaces. (c) Digital twin architecture for evaluation.
    }
    \vspace{-0.2in}
\end{figure*}

As
5G technology being rolled out over the past few years \cite{10.1145/3544216.3544217,10.1145/3544216.3544219, 10723502},
academia and industry have started to envision techniques beyond 5G or 6G \cite{chowdhury20206g, meng2024computer,jiang2021road}.
Ideally, future Radio Access Networks (RAN) will need to support Real-Time (RT) or near-RT intelligent network optimization; however, the current network architectures are not flexible enough to support such functionality. To suffice the requirements, Open RAN (O-RAN)~\cite{oranwhite,polese2023understanding}
has emerged as a promising solution, enabling intelligent, closed-loop control through microservice applications called xApps, ranging from Traffic Steering~(TS) to network slicing deployed on near-RT or non-RT RAN Intelligent Controllers (RICs). This paper focuses on TS, aiming at improving the network spectral efficiency by intelligently and dynamically controlling the cell-level mobility control setting, as opposed to using static settings as in current networks today. The chosen mobility control parameters affect how the O-RAN Distributed Units (O-DUs) make decisions with regard to User Equipment~(UE) handover~(HO) in connected mode and cell reselection in idle mode. As illustrated in Fig.~\ref{fig:problem_demo}, TS xApps configure the parameters, guiding UE HOs between serving and target cells.

Traditionally, cellular networks manage HOs by having each base station (BS) guide UEs to target cells using RAN parameters set by operators. However, the large number and interrelated nature of these RAN parameters often leads network operators to set the values using rough estimates, with uniform ``golden values" applied across regions. This simplification overlooks regional differences in network performance, such as geography, user distribution, and cell deployment, and fails to account for traffic fluctuations over time. Optimization-based methods are also inadequate as they rely on models that may not accurately capture dynamic traffic patterns, especially when the traffic pattern varies, leading to performance issues. While previous frameworks~\cite{adachi2016q,priscoli2020traffic} have attempted per-UE TS by deploying intelligent controllers on each UE or O-DU, this approach is impractical with current RAN architectures, which cannot scale to such fine-grained control. Therefore, we leave per-UE TS for future exploration.

To leverage O-RAN's near-RT and non-RT control capabilities via the RIC, we explore Reinforcement Learning~(RL) for the TS problem due to RL's ability to learn from real-time non-perfect data. However, several challenges arise. First, the variety and variability of traffic patterns (e.g., packet size, periodicity) across multiple cells and frequency bands make training a robust TS policy difficult, requiring large amounts of well-labeled data, which is not easily obtainable. Second, extending trained models to new traffic patterns is challenging. Adapting ML models to new data often requires extensive retraining, which is resource-intensive. Thus, ensuring model extendibility is critical for applying Machine Learning~(ML)/RL-based TS in real RAN environments.
Although recent advances in large models~\cite{sun2024llm, sun2024retrieval, wu2024new}
have shown promise, they are trained with well-labeled data and human feedback, which are hard to obtain in our application practically. 


In light of the aforementioned challenges, we propose Cascade RL~(CaRL) for TS in O-RAN.
We consider multiple UEs and multiple cells with multiple frequency bands in each cell. Each O-DU will have a set of control knobs that will be tuned by the RL controller in a near-RT manner. The goal of the proposed framework is to achieve dynamic load balancing by slightly sacrificing the Quality of Service~(QoS) for high-quality UEs while greatly increasing the QoS for low-quality users.
To reconcile the difficulty of training the policy in such a complex problem, we propose a novel neural network structure as the RL policy that learns to factorize the state space into smaller sub-spaces and to decompose a large policy model into multiple simpler ones corresponding to each sub-space, with each sub-policy making decisions for its corresponding sub-state space.
Another key reason for the multi-sub-policy design of CaRL is to enable knowledge transfer and reduce human intervention during this process. To deploy CaRL to new areas with different data distributions, we only need to initialize new sub-policies from the already trained ones based on the similarity between the new state space and the old ones.
To train the framework, we adopt the offline-RL training technique, Advantage Weighted Actor-Critic~(AWAC) algorithm~\cite{nair2020awac}.

Our key \textbf{contributions} are summarized as follows.
\begin{itemize}
    \item We propose an RL framework, namely CaRL, for O-RAN TS. The proposed algorithm will be trained offline 
    and then deployed on the near-RT RIC for online inference/fine-tuning as an xApp.
    \item To overcome the practical challenge of large state space in the network, we propose state space factorization and policy decomposition method to partition the large state space by setting a factorizer on top of multiple RL sub-policies. Following this, each RL sub-policy will be trained on a sub-state space to optimize the mapping from the sub-space onto the action space.
    \item We propose a knowledge transfer approach for new data distributions (i.e., traffic patterns), where a new sub-policy is initialized from the existing policies based on the similarity of the distributions. The new sub-policy can be further trained to make decisions for the corresponding sub-space more efficiently than starting from scratch.
    \item To evaluate the efficacy of CaRL, we implemented a data-driven O-RAN RIC digital twin using actual trace data from a tier-1 RAN operator. Additionally, we have conducted \underline{field trials on real-world RANs} to evaluate the performance of CaRL.
\end{itemize}


\section{Related Work} \label{sect:related}
To highlight the differences with prior works, we will provide a review of both RL and non-RL-based TS approaches, as well as the idea of factorization and policy decomposition in RL, in the rest of this section.


\textbf{Non-RL-based Traffic Steering}:
Non-RL-based TS approaches mainly model the problem as an optimization problem or game theory problem without involving ML or RL. Marí-Altozano et al.~\cite{mari2019qoe} proposed a QoE-driven algorithm by adjusting the HO margins among cells to achieve load balancing. A rule-based approach, namely Experience Balancing~(EB), is designed, which leverages the fuzzy logic controllers to control the change of the HO margins. This method can perform well for their specific scenario but could fail for data with different distributions because the rules are hard-coded.
Hui et al.~\cite{hui2016novel} proposed a two-step method, where, in the first step, a greedy Grey Relational Analysis~(GRA) algorithm is proposed to solve the network access selection problem and, in the second step, the problem is modeled as a Bounded Knapsack Problem~(BKP), which is solved by a greedy algorithm.
Although this method can find the solution fast, greedy algorithms cannot always guarantee optimality.

\textbf{RL-based Traffic Steering}:
In RL-based TS frameworks, 
researchers~\cite{adachi2016q, priscoli2020traffic}
proposed Q-learning-based RL solutions for TS, where a Q learning agent is trained and deployed on each UE to choose among the available Wi-Fi and Long Term Evolution~(LTE) Access Points~(APs) with a predefined time interval. These works adopt different formulations of the reward functions focusing on different Key Performance Indicators (KPIs).
Unlike the aforementioned RL-based frameworks, our framework considers the complexity of the state and action spaces by training a sub-policy corresponding to each sub-state space so that when the number of cells increases, CaRL can still be efficiently trained.
Furthermore, these RL-based frameworks cannot be easily extended to work for new areas unless they are trained again from scratch. CaRL solves this problem through its multi-sub-policy structure.

\textbf{Space Factorization and Policy Decomposition in RL}:
In RL, state, action spaces, and policies can be factorized or decomposed into simpler components. Sharma et al.~\cite{sharma2017learning} introduced a factorization technique for compositional discrete action spaces, where each dimension is managed by a deep neural network head. However, this method is limited to discrete action spaces. Mahajan et al.~\cite{mahajan2021reinforcement} used tensor decomposition to factorize action spaces by decomposing the transition matrix and rewards, though their approach lacks flexibility and stability. Wong et al.~\cite{wong2021state} proposed a state space decomposition by factorizing the sparse state transition matrix, training separate Q networks for each sub-state space, and combining them via a mixing network. While factorization is common in Multi-Agent Reinforcement Learning (MARL)~\cite{zhou2019factorized}, its relevance is limited in TS problems in O-RAN due to full observability among O-DUs, unlike MARL’s partial observability~\cite{10161019}. Therefore, we will not delve deeply into factorization in MARL.



\section{O-RAN Traffic Steering} \label{sect:o-ran}
In this section, we will explain the O-RAN architecture we adopted and the formulation of the TS problem in O-RAN.

\textbf{Overview of O-RAN Architecture}:
The Open RAN (aka O-RAN) architecture paradigm is realized via the decoupling of the control plane from the user plane in the RAN protocol stack, allowing for innovation of the control plane to happen at a faster pace and fuel the use of AI/ML techniques that leverage near-real-time fine-grained RAN data for more effective RAN management, automation, and optimization. 
The O-RAN architecture mainly consists of three components: i) the RIC platform, ii) the open interfaces connecting the RAN to the RIC to facilitate the RAN data collection and RIC control loops, and iii) the RIC applications (often referred to as the xApps for near-real-time RIC and rApps for non-real-time RIC) that perform RAN data analytics and make intelligent control decisions that can be applied to the RAN via the O-RAN standardized E2, A1, and O1 interfaces.


\textbf{RAN Control Parameters}:
The TS optimization contains two major network features: Connected-Mode Load Balancing (CMLB) and Idle-Mode Load Balancing (IMLB).
For CMLB, it has five control knobs, as follows,
i)~\textit{offloadAllowed}: the indication of whether offloading traffic from the serving cell to the target cell is allowed;
ii)~\textit{maxRC}: the maximum
Residual Capacity~(RC) threshold percentage of the source cells. CMLB can only be triggered on a source cell when its RC is smaller than \textit{maxRC}. A larger value of \textit{maxRC} allows UEs to be offloaded from a source cell more easily;
iii)~\textit{RCHeadroom}:
the RC headroom percentage of target cells. CMLB can be triggered on a target cell only when its RC is larger than \textit{RCHeadroom}. A smaller value of \textit{RCHeadroom} allows UEs to be offloaded to a target cell more easily;
iv)~\textit{deltaRC}: the RC difference between the serving cell and the target cell; and v)~\textit{rsrpCMLBFilter}: the Reference Signal Received Power~(RSRP) threshold of enabling CMLB on a target cell for a given UE. If a UE’s RSRP on a cell is less than \textit{rsrpCMLBFilter}, then this cell will not be considered as a CMLB candidate target cell. IMLB manages the cell selection/reselection for UEs in Radio Resource Control~(RRC)-Idle mode and operates based on a set of weights for each cell. In our framework, the weights are calculated based on the load of the cells; in other words, if the cell is more load congested, it will get a lower weight, and RRC-Idle mode UEs have a smaller chance to camp on this cell and vice versa.

In HO A5 event trigger, we consider one HO offset parameter \textit{cio}. A HO based on an A5 event has to meet two inequalities $RSRP_{serv} + hysThres3InterFreq < thres3InterFreq,$
and
$RSRP_{target} - hysThres3InterFreq + cio > thres3aInterFreq$,
where $RSRP_{serv}$ is the UE's serving cell RSRP, $RSRP_{target}$ is the UE's target cell RSRP. $hysThres3InterFreq$ is a hysteresis to prevent HO ping-pong. $thres3InterFreq$ and $thres3aInterFreq$ are two thresholds applied to serving cell RSRP and target cell RSRP respectively. $cio$ is an offset between a specific serving cell and target cell pair.
By adjusting the values of these knobs, we can control the HO among cells to achieve load balancing. However, the number of these knobs increases as the number of neighboring cells increases. For example, the number of \textit{cio} knobs increases exponentially with the number of neighboring cells because it is defined in a cell-to-cell manner. Therefore, traditional non-RL-based algorithms could easily fail because of a lack of flexibility and scalability.


\vspace{-0.03in}

\section{Proposed Cascade RL Framework} \label{sect:propose}
In this section, we will present in detail our proposed CaRL-based TS framework.
We will first introduce the background in Sect.~\ref{sect:propose:background}.
Then, in Sect.~\ref{sect:propose:splitting}, we will present the proposed CaRL framework, the offline training approach, and the knowledge-transferring technique for CaRL.

\vspace{-0.1in}
\subsection{Background and Notations} \label{sect:propose:background}

Problems that can be modeled as a Markov Decision Process~(MDP) can be solved by RL algorithms. An MDP consists of a state space $\mathcal{S}$, which characterizes the properties of the system, and an action space $\mathcal{A}$. Furthermore, the core of an RL agent is the policy $\pol: \mathcal{S} \rightarrow \mathcal{A}$ with parameters $\theta$, which make decisions based on the incoming state. In this paper, we use $\pol$ and $\cpol$ to represent stochastic and deterministic policies, respectively. After executing an action, the environment will change according to a possibly unknown state transition function $\mathcal{T}: \mathcal{S} \times \mathcal{A} \rightarrow \mathcal{S}$. Then the environment will generate a reward to the agent as a function of state and action $r: \mathcal{S} \times \mathcal{A} \rightarrow \mathbb{R}$. The goal of RL algorithms is to maximize the expected return $R=\sum_{t=0}^T \gamma^t r_t$, where $T$ is the maximum number of steps, and $\gamma$ is the discount factor.

Policy Gradient~(PG) is a popular algorithm to solve the MDP by directly calculating the gradient of the policy function and optimizing it towards the optimal direction. 
The gradients of the policy mentioned above can be written as,
$ J(\theta) = \mathbb{E}_{s \sim p^{\pol}, a \sim \pol_{\theta}} [R]$, where $p^{\pol}$ is the state transition probability.
In addition, to handle MDPs with continuous action spaces,
Deterministic Policy Gradient~(DPG)~\cite{silver2014deterministic} and DDPG~\cite{lillicrapHPHETS15} are proposed, with the latter being the neural-network version, and the policy gradient can be written as,
\begin{align} \label{eq:ddpg}
    \nabla_\theta J(\theta) = \mathbb{E}_{s \sim \mathcal{D}} \left[ \nabla_\theta \cpol (s) \nabla_a Q^{\cpol} (s, a) |_{a=\cpol (s)} \right],
\end{align}
where $\cpol$ is the deterministic policy, and $\mathcal{D}$ represents the Experience Replay (ER) buffer which contains tuples $<s, a, r, s'>$.
DDPG also utilizes a Q network (also known as the critic), trained by reducing the estimation error between it and the real returns, $\mathcal{L}(\theta)=\mathbb{E}_{s,a,r,s'} [(Q^{\cpol} (s, a) - y)^2]$,
where $y=r + \gamma Q^{\cpol'}(s', \cpol'(s'))$, with $\cpol'$ being the target policy network with delayed parameters to stabilize training.


\subsection{Cascade Reinforcement Learning} \label{sect:propose:splitting}
We will present CaRL from three angles--state space factorization and policy decomposition,
offline training,
and knowledge transferring.

\textbf{State Space Factorization and Policy Decomposition:}
In RL, the agent essentially tries to optimize the mapping from the state space onto the action space by interacting with the environment. However, if the state and action spaces are too large and complex, even with the help of Deep NN~(DNN), an agent might still fail because complex DNNs are hard to be trained.
In TS in O-RAN, the state space consists of important features to describe the cell, including the time of day, day of the week, cell Physical Resource Block~(PRB) utilization, and so on, and the action space consists of multiple control knobs to adjust the HO margin. Although the state and action spaces do not seem to be large, that is only for one cell, and we usually consider more than 10 cells in reality. Therefore, to overcome this difficulty, we propose a cascade RL architecture to divide and conquer the problem.
We consider state space factorization to mitigate the complexity of the state space. Specifically, we train a factorizer that tries to classify the incoming state into multiple sub-spaces. The factorizer can be any model, including DNN, 
decision tree,
and so on, as long as the output of it is a probability distribution over the sub-spaces. After the state is factorized into one of the sub-spaces, the corresponding sub-policy will then make decisions.
The structure of the cascade policy is shown in Fig.~\ref{fig:CaRL}.
\textit{To avoid any confusion on the terms we use, we refer to the whole RL model, including the factorizer and the bottom-level models as the RL policy and the bottom RL models as sub-policies}.

The factorizer can be regarded as a state space splitter, and it tries to determine which sub-space the current incoming state belongs to. To illustrate, suppose we have a factorizer $C: \mathcal{S} \rightarrow Pr(\Cpol)$ with parameters represented by $\theta_{C}$, where $\Cpol=\{\cpol_1,..., \cpol_M\}$ is the set of sub-policies corresponding to $M$ sub-state spaces. The factorizer tries to assign the incoming state to one of the policies by generating a probability distribution over the $M$ sub-spaces. Note that since the action spaces are continuous, we are adopting deterministic policies $\cpol_i$ for sub-space $i$. Generally, traditional classifiers are trained with labeled data; however, we do not have data with accurate labels for sub-spaces. Although unsupervised learning algorithms such as clustering methods can be used as the factorizer in our framework, however, most of these algorithms assign the same importance to every feature, which could be misleading. As such, to train the factorizer, we propose to incorporate the factorizer into the policy gradient paradigm and directly optimize it by applying the chain rule. That is, the maximization of the accumulated expected reward is regarded as the downstream task for the factorizer. 

As for the bottom part of the model, we have $M$ RL sub-policies that correspond to $M$ sub-spaces. $M$ is a predefined hyperparameter and can have greater values if the problem is more complicated and vice versa. When $M$ is equal to $1$, the framework boils down to the DDPG algorithm~\cite{lillicrapHPHETS15}, where we consider the whole state space at the same time. Each sub-policy is represented by a neural network $\cpol_i$ with parameters $\theta_i$.
In the proposed framework, we assume that for a state $s \in \mathcal{S}$, if it is in the center area of sub-space $\mathcal{S}_i$, then it should be assigned to sub-policy $\cpol_i$. However, if $s$ resides at the border of $\mathcal{S}_i$, then the decision should be made by the multiple policies that are closest to the state. This process is realized by letting the factorizer generate a soft classification score, which is a probability distribution $\classifier(s)=\{c_1, ..., c_M\}^T,\ \sum_{i=1}^M c_i = 1$ over the sub-spaces and this probability distribution will be used as the weight to calculate the weighted average on the actions generated by each sub-policy. Therefore, the final action can be calculated by,
$a=\Cpol(s)^T \cdot \classifier(s)=\sum_{i=1}^M c_i \cpol_i(s),$
where $\Cpol(s)=\{\cpol(s)_1, ..., \cpol(s)_M\}^T$ is the set of outputs from the sub-policies given the state $s$.
However, a hard factorizer that generates binary classification results can also be trained with the help of the Gumbel Softmax trick~\cite{jang2016categorical}, which approximates the sampling operation from categorical distributions and allows the gradient to flow through it.


\textbf{Offline Training:}
Since directly training the joint policy online in the real RAN from scratch can be risky,
we train CaRL offline with a pre-collected dataset from a tier-1 RAN operator.
This dataset consists of MDP data (i.e., state and corresponding action and reward) collected by applying a conservative heuristic algorithm (refer to Sect.~\ref{sect:eval:data_sim}) to the RAN. However, offline RL training is essentially different from traditional supervised learning, as the dataset does not necessarily contain optimal actions.
Therefore, the supervised learning technique does not suffice in the RL scenario. A more important problem in offline training is that, given the state, the corresponding action and reward sometimes do not exist in the data set for the RL policy.
Some researchers use bootstrap techniques to estimate reward values for non-existing actions, but it can lead to significant deviation in the estimation of the Q value, also known as the \textit{bootstrap error} problem~\cite{sutton2018reinforcement}.
To solve the aforementioned problems, we adopt an improved AWAC~\cite{nair2020awac}, an offline RL algorithm that imposes a constraint on the optimization of the policy to limit the generation of actions that are not contained in the dataset.

The optimization of the policy in AWAC is derived from~\eqref{eq:ddpg}.
It adds a KL-divergence constraint to limit the difference between the policy being trained and the policy used to collect data, and a closed-form solution can be obtained by applying the KKT condition~\cite{peng2019advantage}.
The policy loss for AWAC can be represented as,
\begin{align} \label{eq:awac_pg}
    J(\theta) = -\mathbb{E}_{s, a \sim \beta} \left[ \log \cpol (s) \exp\left(\frac{1}{\lambda} A^{\cpol} (s, a)\right)
    \right],
\end{align}
where $\lambda$ is the Lagrangian multiplier, and $\beta$ is the offline dataset collected before training.

In addition to policy loss, we encourage the factorizer to generate confident output distributions, which means that the factorizer output distribution should be close to the one-hot distribution. The reason is that if the output distribution is close to the other extreme, i.e., the uniform distribution, state space factorization is meaningless because the sub-policies have the same weights for every incoming state. As such, we add a negative entropy regularization term to the output of the factorizer.
The loss for the offline training of CaRL, $J(\boldsymbol{\theta})$, can be written as,
\begin{align} \label{eq:carl_pg}
    -\mathbb{E}_{s, a \sim \beta} \left[ \log \left( \mathcal{C}(s)^T \cdot \Cpol(s) \right) \exp\left(\frac{1}{\lambda} A^{\cpol} (s, a)\right)
    \right] - \sum_{i=1}^M c_i ^2,
\end{align}
where $\boldsymbol{\theta}=\{ \theta_C, \theta_1, \theta_2, ..., \theta_M \}$ represents the parameters for the factorizer and the sub-policies, and $\mathcal{H}(C(s))$ is the entropy regularization term. Now that the action is calculated by combining the factorizer's output and the sub-policies outputs, the parameters of the factorizer $\theta_C$ can be updated according to~\eqref{eq:carl_pg} via backpropagation. The policy gradient of~\eqref{eq:carl_pg} w.r.t. the $i$th sub-policy, $\nabla_{\theta_i}J(\boldsymbol{\theta})$, can be written as,
\begin{align} \label{eq:grad_subpol}
    \mathbb{E}_{s, a \sim \beta} \left[ \frac{c_i \nabla_{\theta_i} \cpol_i(s)}{\sum_{i=1}^M c_i \cpol_i(s)} \exp\left(\frac{1}{\lambda} A^{\cpol} (s, a)\right)
    \right].
\end{align}
The policy gradient of \eqref{eq:carl_pg} w.r.t. the factorizer, $\nabla_{\theta_C}J(\boldsymbol{\theta})$, can be written as,
\begin{align} \label{eq:grad_classifier}
    \mathbb{E}_{s, a \sim \beta} \left[ \frac{\Cpol(s)^T \nabla_{\theta_\classifier} \classifier(s)}{\classifier(s)^T \Cpol(s)} \exp\left(\frac{1}{\lambda} A^{\cpol} (s, a)\right)
    \right] - \nabla_{\theta_C}\mathcal{H}(C(s)).
\end{align}
In~\eqref{eq:grad_classifier}, we can clearly see that the factorizer gradient can be obtained by taking the derivative of the downstream loss function.
In addition to the policy, we use two Q networks to determine whether an action is good or not, following soft actor critic~(SAC)~\cite{haarnoja2018soft} to avoid an overestimation of the Q value problem that could occur during training. Since we have not changed the Q networks update method, we omit further technical details due to the space limit.


\textbf{Knowledge Transferring:}
Another benefit of state space factorization is that the trained policy can be easily adapted to new traffic patterns, \emph{which is especially needed by RAN operators when applying trained policies to new network areas, e.g., adapting policy trained on urban areas to rural areas.}
To achieve this, we propose a knowledge-transfer approach based on CaRL architecture, which can greatly reduce
the time consumed by retraining a model from scratch \cite{zhuang2020comprehensive}.
To enable knowledge transferring from a trained CaRL model to new traffic patterns, we add new sub-state spaces and initialize new corresponding sub-policies. However, the new sub-policies do not start from scratch with zero knowledge.
Taking advantage of the soft probability distribution generated by the factorizer, we obtain the actions for the new sub-policy by calculating the weighted average of the parameters of the trained sub-policies using the probability distribution as weights.
Suppose that the states from new traffic patterns are represented by $\boldsymbol{s}^{new}=\{ s^{new}_i, ..., s^{new}_K \}$, where $K$ is the number of samples in the dataset. The initial parameters for the new sub-policy, $\theta_{new}$, can be calculated as,
\begin{equation} \label{eq:knowledge_transfer}
    \theta_{new} = \frac{1}{K}\sum_{j=1}^K C(s^{new}_j)^T \cdot \boldsymbol{\theta},
\end{equation}
where $\frac{1}{K}\sum_{j=1}^K C(s^{new}_j)$ is the normalized weights obtained by averaging the outputs from the factorizer using the new data as input.

\begin{figure}[t!]
    \centering 
    \begin{subfigure}{0.24\textwidth}
      \includegraphics[width=\linewidth]{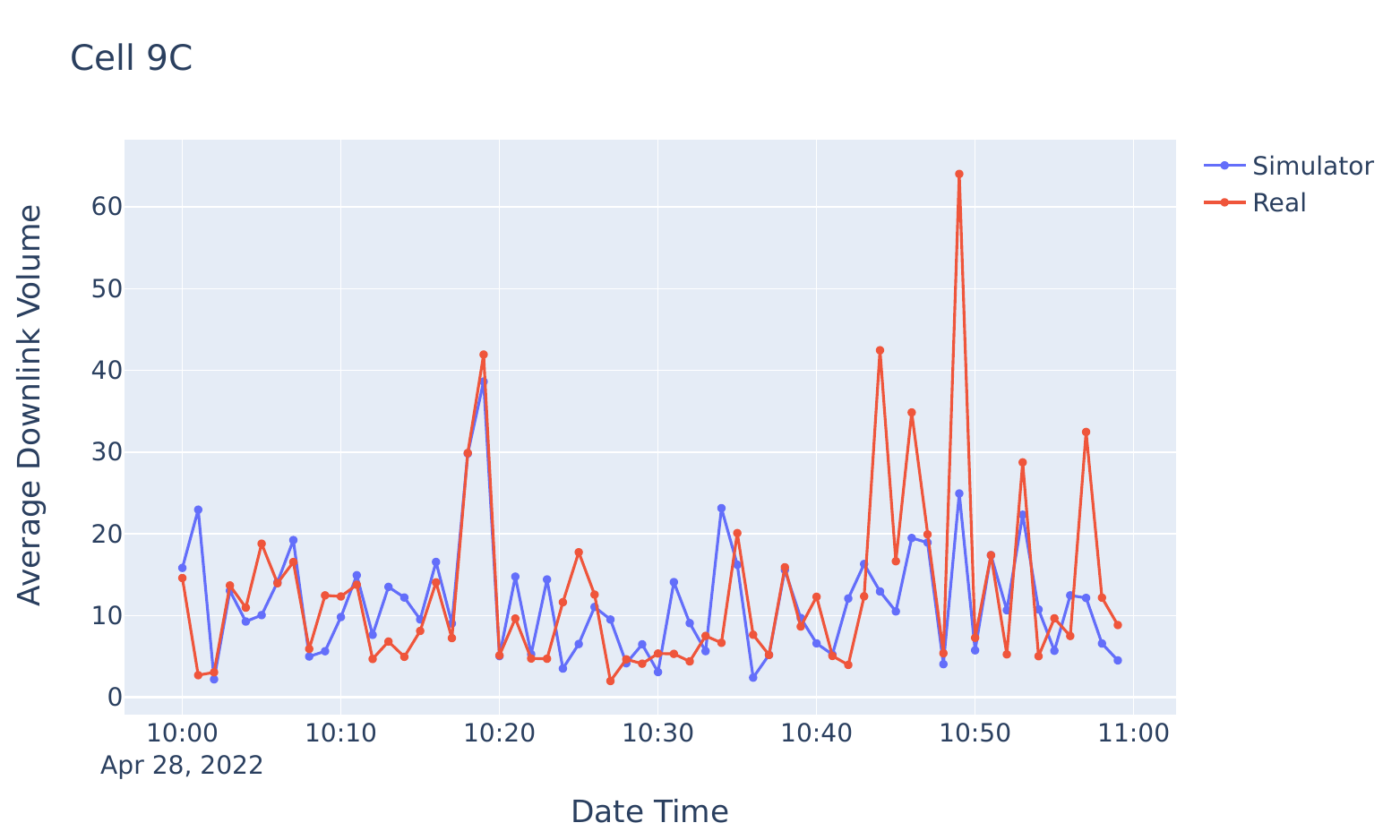}
      \caption{\label{fig:stem_vs_sim:vol}}
    \end{subfigure} 
    \begin{subfigure}{0.24\textwidth}
      \includegraphics[width=\linewidth]{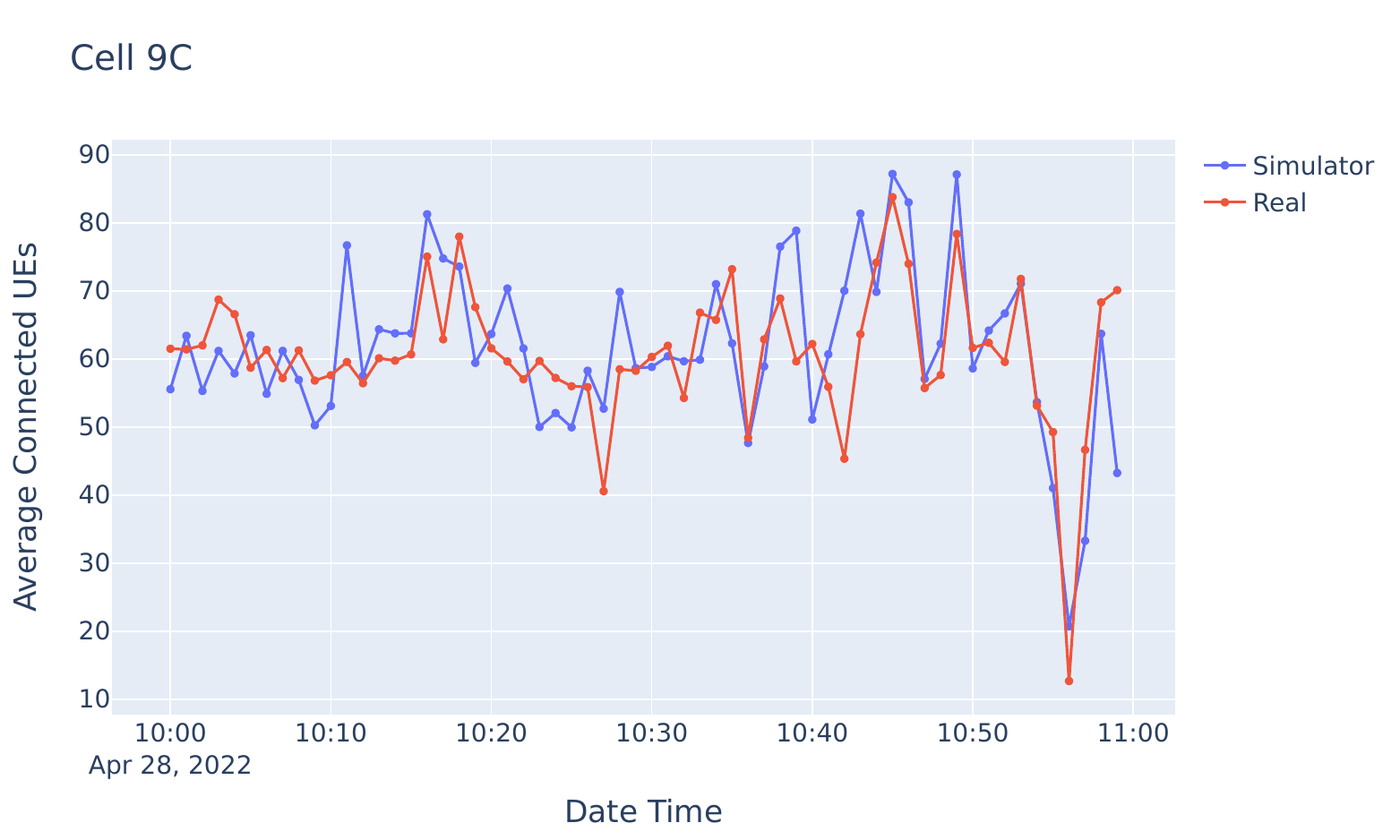}
      \caption{\label{fig:stem_vs_sim:nue}}
      \label{fig:steps-vs-agents}
    \end{subfigure} 
    \caption{\label{fig:stem_vs_sim} Comparison between the real-world network trace data and the digital twin data for a specific cell in cluster 2
    over a one-hour period for (a) sum downlink volume and (b) average number of RRC connected UEs. Due to proprietary restrictions, we cannot disclose the actual numbers of the volume, throughput, and the number of handovers. Instead, we normalized the values in this figure and provided the percentage. 
    }
    \vspace{-0.2in}
\end{figure}

\begin{figure*}[t!]
    \centering 
    \begin{subfigure}{0.3\textwidth}
      \includegraphics[width=\linewidth]{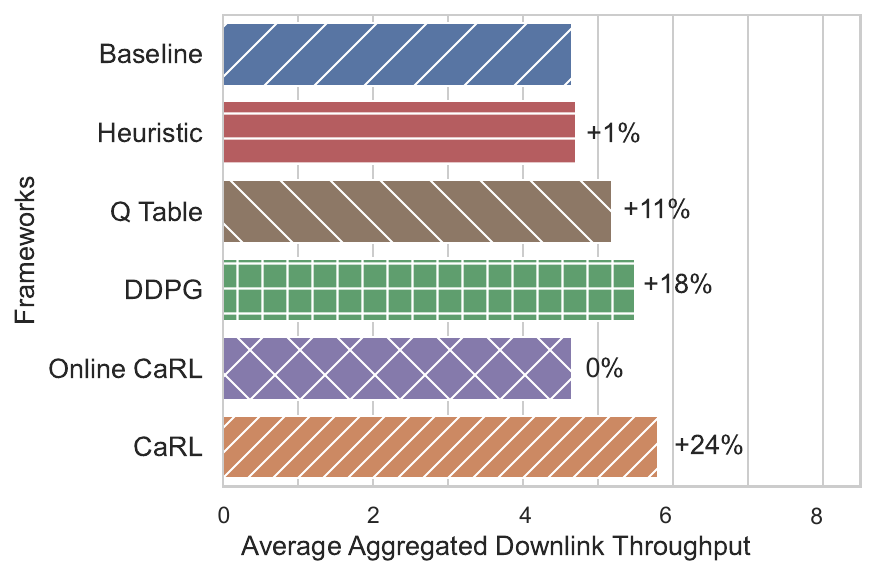}
      \caption{\label{fig:avg_tput_somerville} Avg. aggregated downlink throughput.}
    \end{subfigure}\hfil 
    \begin{subfigure}{0.3\textwidth}
      \includegraphics[width=\linewidth]{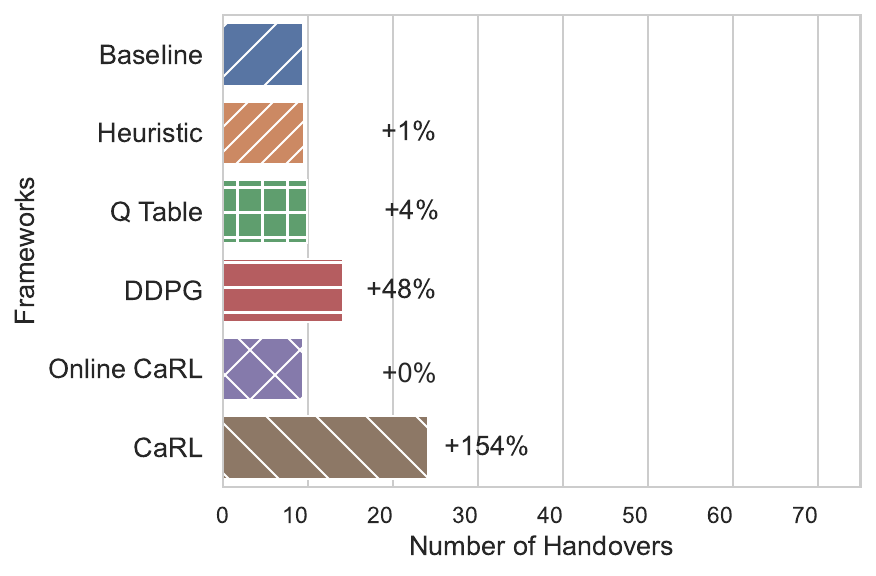}
      \caption{\label{fig:total_ho_somerville} Total number of Handovers.}
    \end{subfigure}\hfil 
    \begin{subfigure}{0.3\textwidth}
      \includegraphics[width=\linewidth]{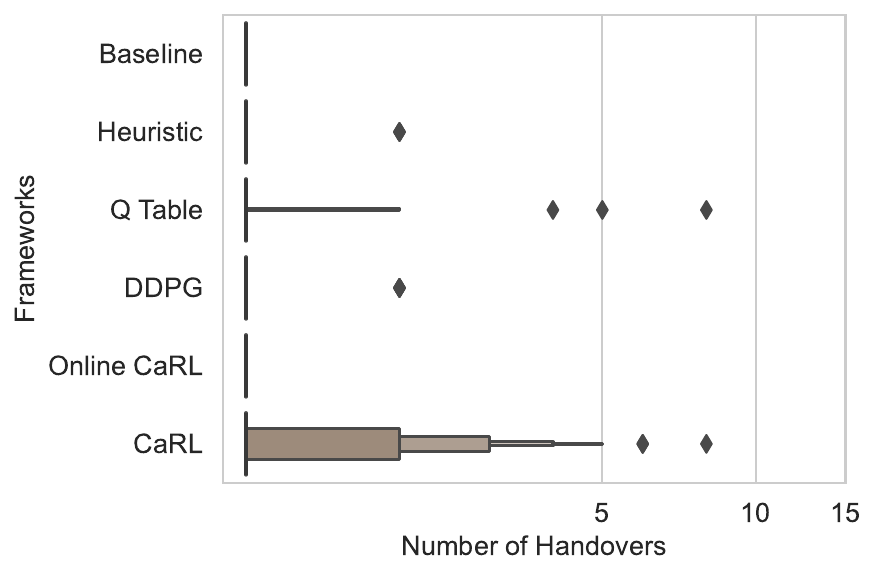}
      \caption{\label{fig:ho_somerville} Distribution of \#HOs per UE.}
    \end{subfigure}\hfil 
    \caption{\label{fig:sim_somerville} Simulation results generated using RAN configuration data and traffic data from cluster 1. Due to proprietary restrictions, we cannot disclose the actual numbers of the volume, throughput, and the number of handovers. Instead, we normalized the values in this figure and provided the percentage.}
    \vspace{-0.1in}
\end{figure*}

After obtaining the new sub-policy, the joint policy still needs to be fine-tuned on new traffic data to achieve the best performance. In this way, instead of initializing a blank model,
the network operators can feed new MDP data to the extended models for a few rounds of fine-tuning and then deploy it into the system. Or, they can fine-tune the new policy even at run-time but with a much shorter training time. 
In our evaluation, to demonstrate the performance of knowledge transferring, we do not fine-tune the new sub-policy but directly evaluate it, i.e., the direct results of \eqref{eq:knowledge_transfer}, on the new traffic data.

\begin{figure*}[t!]
    \centering 
    \begin{subfigure}{0.3\textwidth}
      \includegraphics[width=\linewidth]{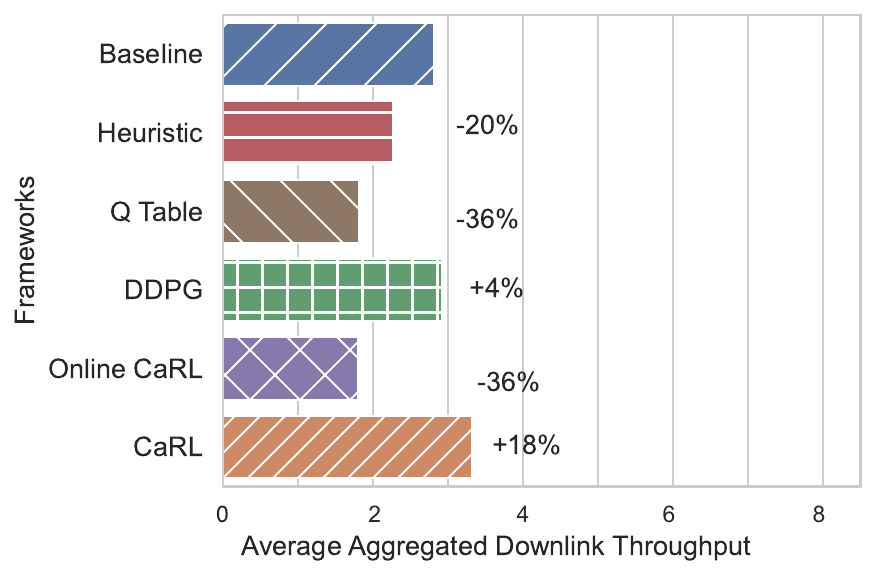}
      \caption{\label{fig:avg_tput_ohio} Avg. aggregated downlink throughput.}
    \end{subfigure}\hfil 
    \begin{subfigure}{0.3\textwidth}
      \includegraphics[width=\linewidth]{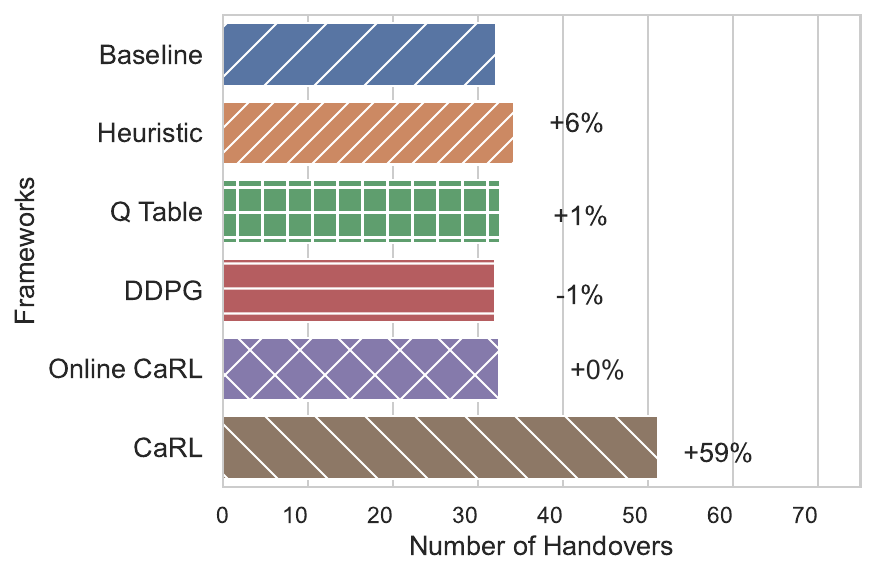}
      \caption{\label{fig:total_ho_ohio} Total number of Handovers.}
    \end{subfigure}\hfil 
    \begin{subfigure}{0.3\textwidth}
      \includegraphics[width=\linewidth]{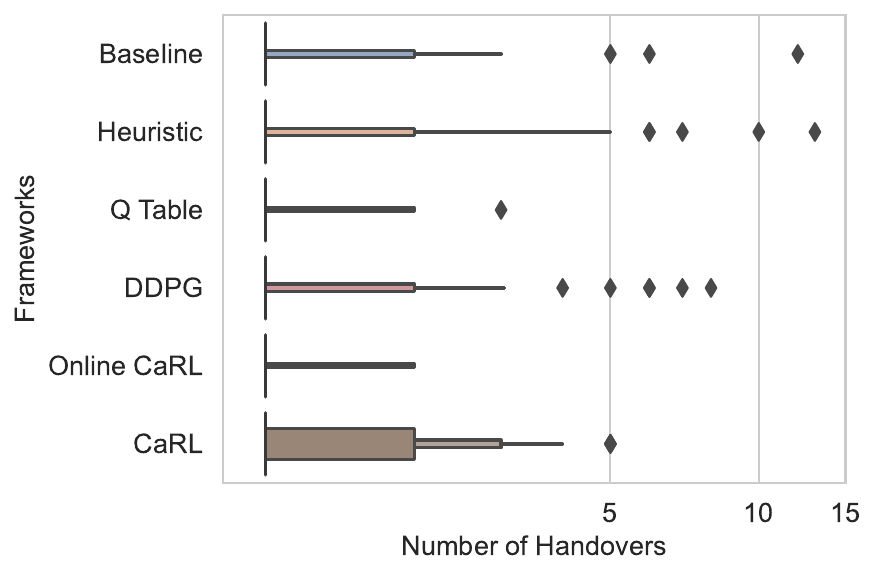}
      \caption{\label{fig:ho_ohio} Distribution of \#HOs per UE.}
    \end{subfigure}\hfil 
    \caption{\label{fig:sim_ohio} Simulation results generated using RAN configuration data and traffic data from cluster 2. Due to proprietary restrictions, we cannot disclose the actual numbers of the volume, throughput, and the number of handovers. Instead, we normalized the values in this figure and provided the percentage.
    }
    \vspace{-0.2in}
\end{figure*}

\subsection{MDP Formulation}
To solve the TS problem with RL, we need to model it as an MDP consisting of state, action, and reward. The state in MDP should capture all the information needed for the optimization.
We formulate the state and action in a cell-to-cell manner, i.e., serving/target cell pairs, by including information about both the serving cell and the target cell.
In our problem, the \underline{state} is defined as $s=$[\textit{serving cell frequency, target cell frequency, time of day, day of the week, serving cell PRB utilization, target cell PRB utilization, serving cell throughput, target cell throughput}]. Due to the complexity and scale of this state space, especially when the number of cells in the target areas is large, the problem becomes hard to solve.
As for the \underline{action},
we use the six control knobs introduced in Sect. \ref{sect:o-ran}.
For \underline{reward},
we use the cell-level average aggregated downlink throughput, which can be calculated by
$throughput\ (Mbps)=downlink\ volume\ (Mbits) / transmission\ time\ (s)$.

\section{Performance Evaluation} \label{sect:eval}

In this section, we present the evaluation setup and results obtained from the digital twin~(Sect.~\ref{sect:eval:data_sim}) and field trials in real-world RAN (Sect.~\ref{sect:eval:field}).



\subsection{Digital Twin Evaluation}\label{sect:eval:data_sim}
\textbf{Digital Twin Framework}:
Assessing the effectiveness of RIC applications remains a challenging task despite its many potential benefits. This is due to the lack of fully-compliant O-RAN implementations, both commercially and experimentally, for measurement reporting and control service models. Additionally, the current realization of O-RAN is still at a very limited scale, mostly consisting of a few BSs in laboratory or commercial trial environments, making it difficult to fully evaluate the value of O-RAN data feeds and control loops and their impact on large-scale networks.
Open-source RAN software frameworks and testbeds, such as COSMOS~\cite{raychaudhuri2020challenge}, POWDER~\cite{breen2020powder}, srsLTE~\cite{gomez2016srslte}, and OpenAirInterface (OAI) \cite{nikaein2014openairinterface}, can be used as an alternative to commercial-grade networks. These frameworks can be integrated with the O-RAN software platform by connecting to the RIC through the E2 interface. However, there are some significant limitations associated with these systems. Firstly, they have limited scales, supporting only a small number of RAN nodes and UEs. Secondly, the functionality of the RAN nodes is not on par with commercial RAN and does not support advanced mobility and traffic management. Finally, it is challenging to recreate traffic patterns with geographical and temporal variations found in actual networks. Additionally, traditional network simulators, such as ns-3 \cite{riley2010ns}.


\begin{figure*}[t!]
    \centering 
    \begin{subfigure}{0.5\textwidth}
      \includegraphics[width=\linewidth]{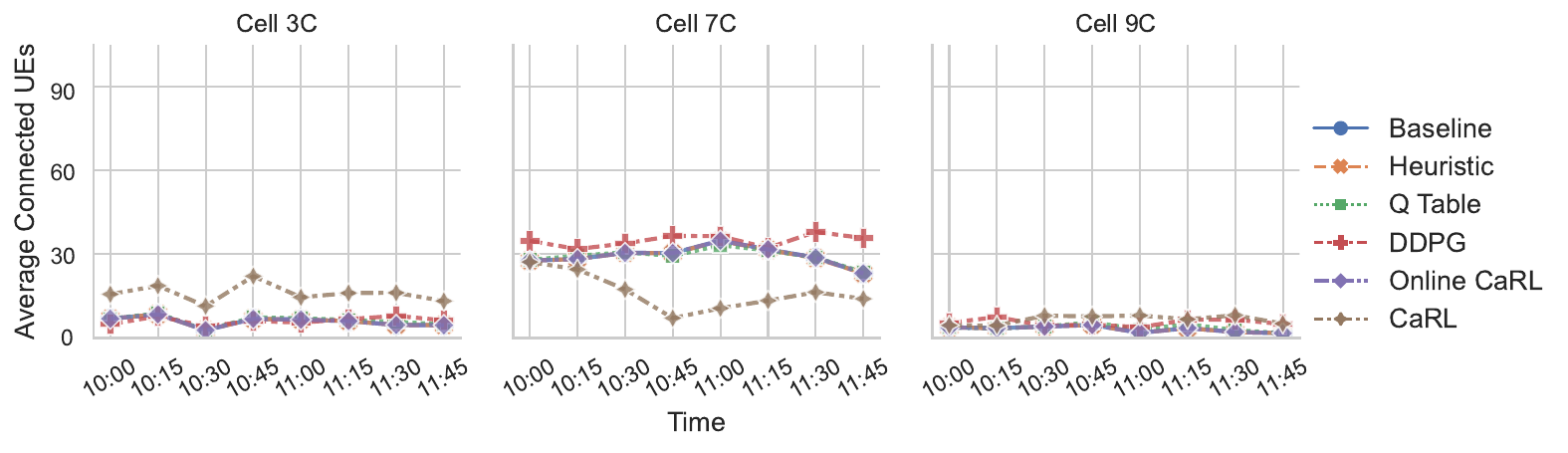}
      \caption{\label{fig:cells_nue_somerville} Three cells in a sector in cluster 1.}
    \end{subfigure}\hfil 
    \begin{subfigure}{0.5\textwidth}
      \includegraphics[width=\linewidth]{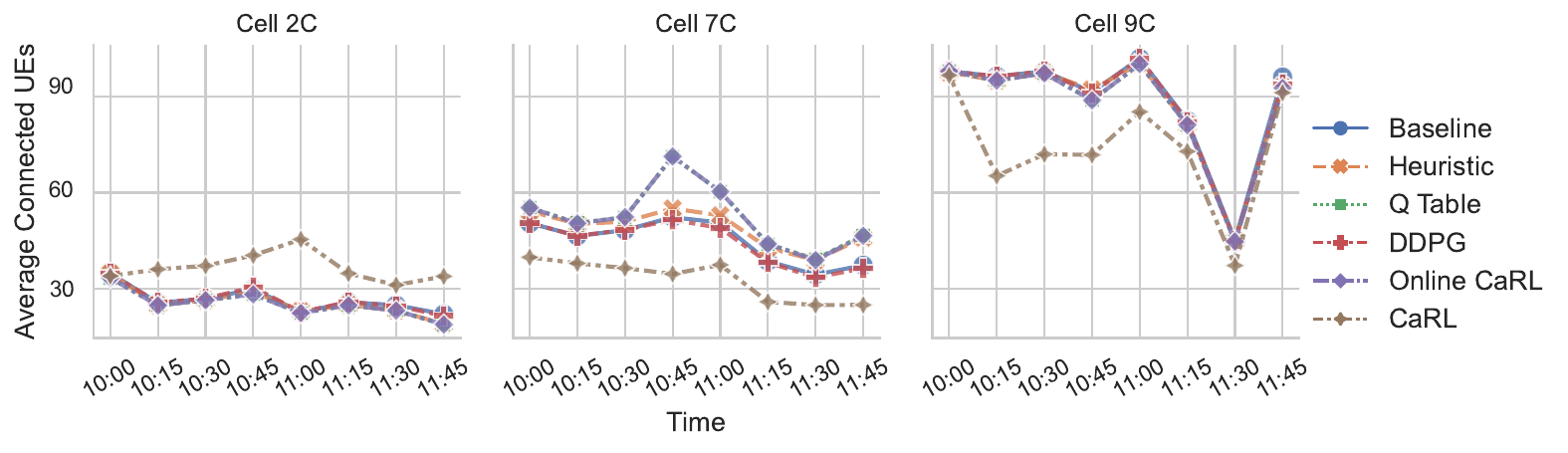}
      \caption{\label{fig:cells_nue_ohio} Three cells in a sector in cluster 2.}
    \end{subfigure}\hfil 
    \caption{\label{fig:cells_nue} Average connected UEs for three co-located cells in the same sector. Due to space limitations, we only present the results for one sector in each cluster.  Due to proprietary restrictions, we cannot disclose the actual numbers of the volume, throughput, and the number of handovers. Instead, we normalized the values in this figure and provided the percentage. 
    }
    \vspace{-0.1in}
\end{figure*}

\begin{figure*}[t!]
    \centering 
    \begin{subfigure}{0.5\textwidth}
      \includegraphics[width=\linewidth]{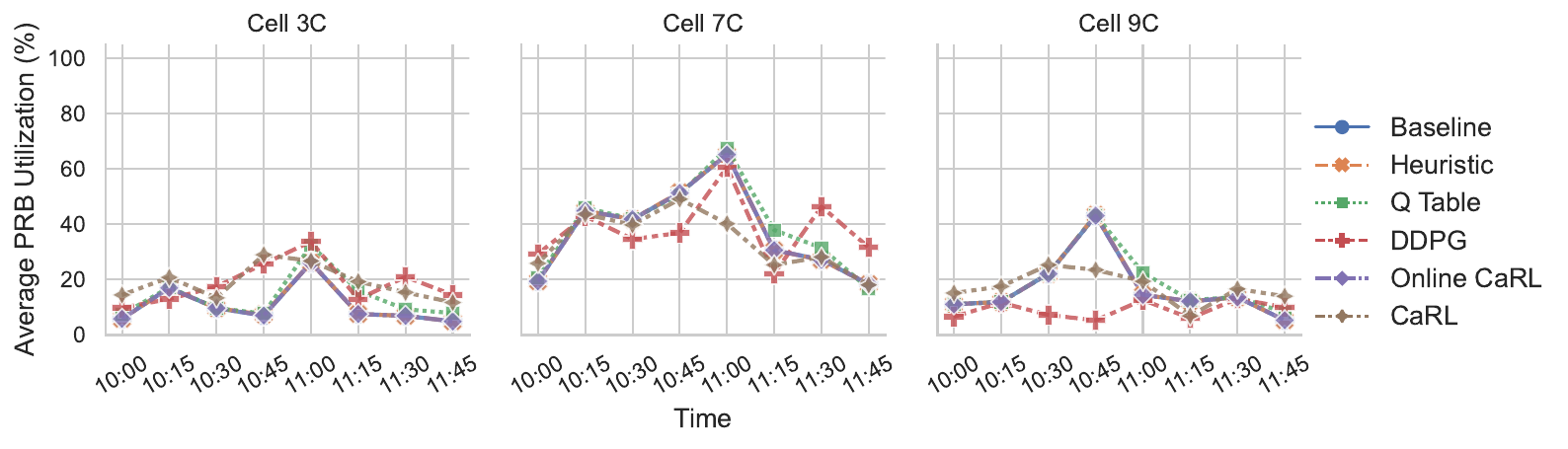}
      \caption{\label{fig:cells_prb_somerville} Three cells in a sector in cluster 1.}
    \end{subfigure}\hfil 
    \begin{subfigure}{0.5\textwidth}
      \includegraphics[width=\linewidth]{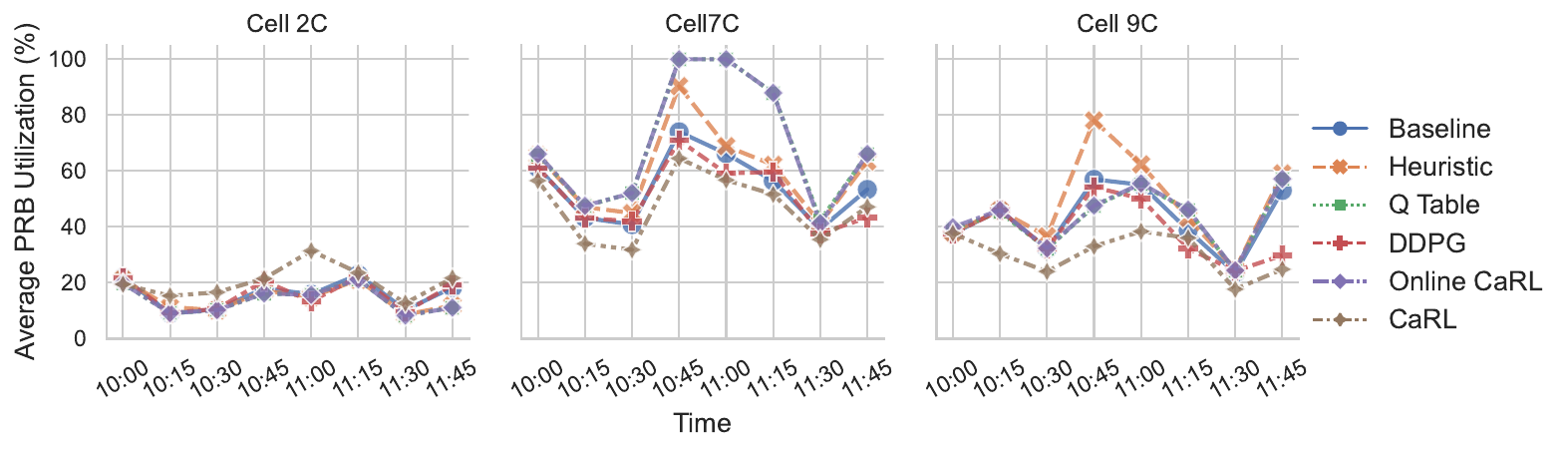}
      \caption{\label{fig:cells_prb_ohio} Three cells in a sector in cluster 2.}
    \end{subfigure}\hfil 
    \caption{\label{fig:cells_prb} PRB utilization for three cells co-located in the same sector. Due to proprietary restrictions, we cannot disclose the actual numbers of the volume, throughput, and the number of handovers. Instead, we normalized the values in this figure and provided the percentage. 
    }
    \vspace{-0.2in}
\end{figure*}

To overcome the limitations of current O-RAN testbeds and simulation/emulation frameworks, we developed a data-driven and scalable network digital twin framework that aims to realistically model actual cellular network deployment and behavior (Fig.~\ref{fig:sim}). This is made possible via the use of real-world data from a tier-1 RAN operator to create digital twin scenarios that reflect the actual cell deployment topology, carrier frequency, and antenna configurations. 
Our RAN protocol stack features a PHY-layer abstraction model and a detailed MAC-layer scheduling with per-Transmission Time Interval~(TTI) and per-QoS Class Identifier~(QCI) level PRB allocation, as well as control logics at RLC and RRC layers. Additionally, we integrated advanced mobility control functionalities that mimic the operator's policies, such as idle-mode cell reselection, intra-frequency and inter-frequency HOs, and load balancing, all of which are crucial in establishing a representative baseline for testing our advanced traffic steering algorithm. One of the important components in our digital twin framework is the controller interface that enables the implementation of simulated RIC applications, providing the ability to read fine-grained RAN state information and make control actions in a similar manner as the O-RAN RIC. 

Matching network KPIs obtained from simulations with reported KPIs of the RAN is challenging overall. This difficulty arises from several factors, including missing modeling inputs like actual user positions and mobility patterns, as well as the unpredictable nature of the radio environment. However, our proposed solution demonstrates promising results, as depicted in Fig.~\ref{fig:stem_vs_sim}, where we compare the behavior of our simulator to actual RAN measurements on two selected KPIs. The comparison reveals that our simulated KPIs closely follow the overall trend and maintain a reasonably close proximity to the actual KPIs.


\textbf{Dataset}:
The training data consists of state, action, next state, and reward, which are collected using a heuristic algorithm from the BSs in the real world in cluster 1. We use this data to train the CaRL model offline and deploy it in different scenarios in the simulator to demonstrate its performance.
We set up two BS clusters, which are selected from two US cities operated by a tier-1 RAN operator. The two scenarios created from cluster~1 and 2 have $12,950$ and $16,194$ unique UEs and $23,000+$ Gbits and $36,000+$ Gbits of total traffic volume over a 2-hour period, respectively. 
In cluster~1, we have 3 LTE BSs, a total of 12 cells on 3 different frequencies, while in cluster~2, there are 3 LTE BSs, a total of 18 cells on 5 frequencies. Each cell has a bandwidth of 10MHz.


\textbf{Experiment Setup and Comparison Frameworks:} 
The factorizer and the sub-policies are realized as fully connected NNs with Adam optimizer and learning rate of 0.0001. The batch size is set to be 256. We set $M$ as 5 for cluster~1. Then, we apply the knowledge transfer technique for cluster 2. The RAN control parameters were tuned every 15 minutes. In this paper, we utilized the AWAC method to train the model instead of proposing a new RL training scheme. As a result, we do not compare our results with other training schemes.

In the simulation on the digital twin, we only present results for offline-trained CaRL because adding online fine-tuning would make it unclear which part of training contributes to the performance. On the other hand, let us point out that we did combine offline and online fine-tuning in the field trials to adapt to new data (Sect.~\ref{sect:eval:field}).

We compare CaRL with the other five frameworks.
i) \textit{Baseline}: this is the default scenario using mobility control parameters as currently set by the network operator.
ii) \textit{Heuristic Algorithm}: we adjust the parameter values step by step following a set of rules from experiences and heuristics. For example, if cell A is more loaded than cell B, then we should turn \textit{offloadAllowed} on and increase \textit{maxRC} by $2\%$.
iii) \textit{Q-table}: we directly use the offline dataset and rearrange it into a Q-table.
iv) \textit{DDPG}: as mentioned in Sect.~\ref{sect:propose}, when the number of sub-space is equal to one, CaRL reduces to DDPG algorithm. Therefore, to show the advantage of the state space factorization and policy decomposition, we train a DDPG agent with AWAC algorithm offline for comparison.
v) \textit{Online CaRL}: we do not apply offline training and let the RL agent learn from scratch (i.e., cold start).



\begin{figure}[t!]
    \centering
    \includegraphics[width=0.25\textwidth]{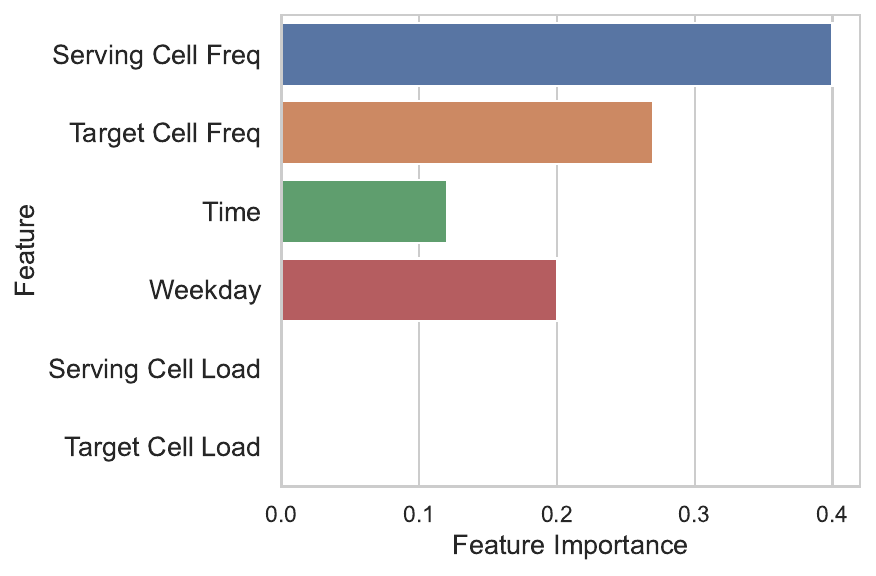}
    \caption{\label{fig:feature_imp} Feature importance obtained by feeding the training data to the trained factorizer to obtain the labels for each state in the dataset. Then the labels will be used to train a decision tree for the feature importance.
    }
    \vspace{-0.22in}
\end{figure}

\begin{figure*}[!t]
    \centering 
    \begin{subfigure}{0.33\textwidth}
      \includegraphics[width=\linewidth]{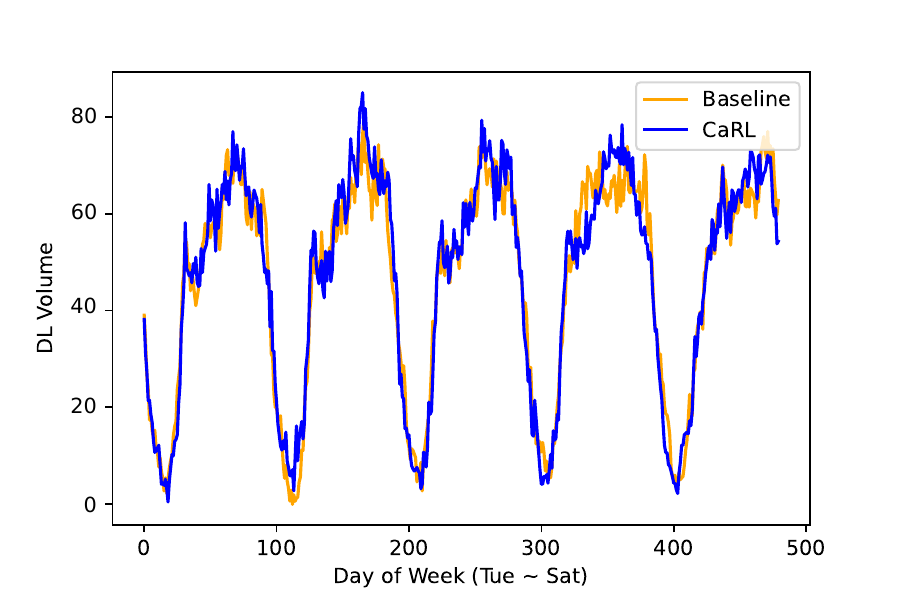}
      \caption{\label{fig:field_vol} Downlink traffic volume.}
    \end{subfigure}\hfil 
    \begin{subfigure}{0.33\textwidth}
      \includegraphics[width=\linewidth]{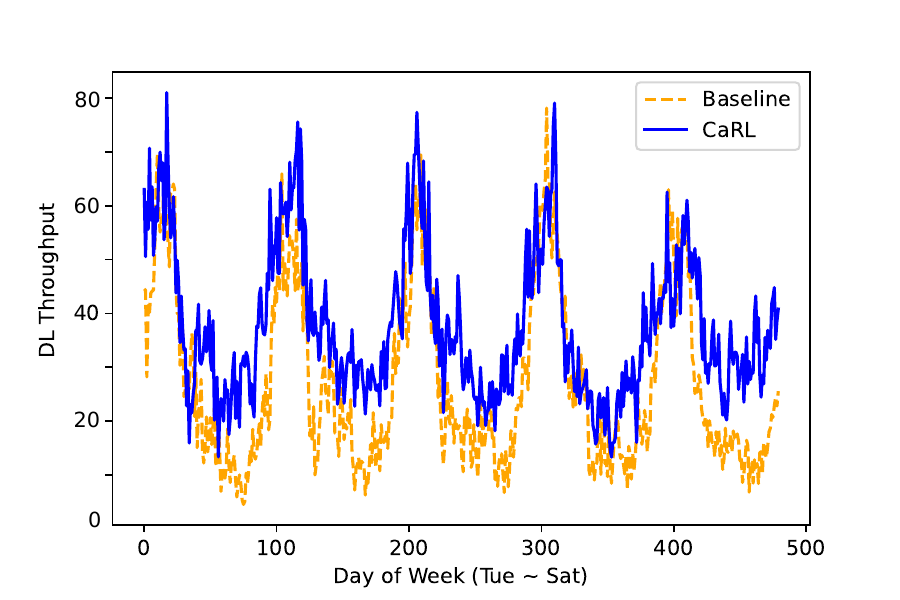}
      \caption{\label{fig:field_thrpt} Downlink throughput.}
    \end{subfigure}\hfil 
    \begin{subfigure}{0.33\textwidth}
      \includegraphics[width=\linewidth]{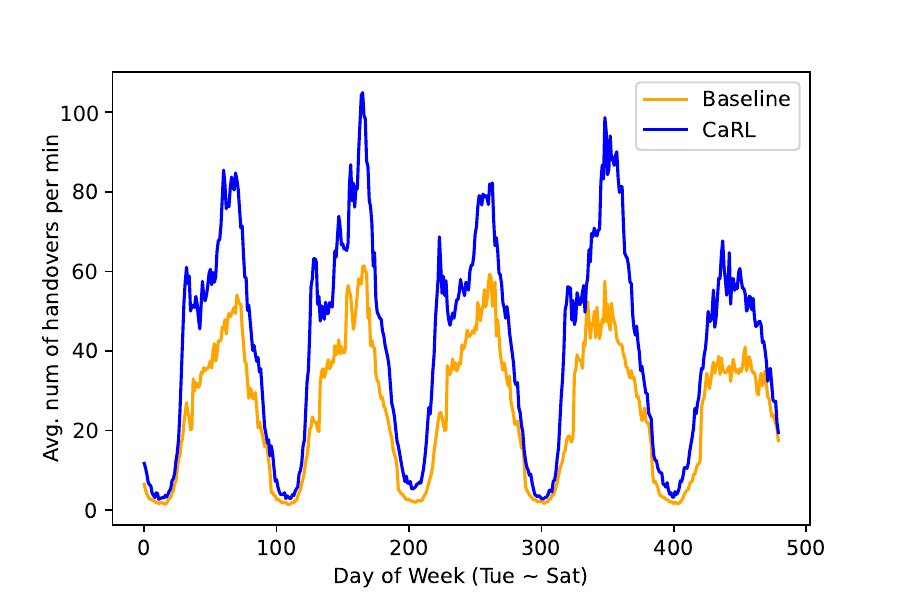}
      \caption{\label{fig:field_ho} Average number of HOs per minute.}
    \end{subfigure}\hfil 
    \caption{\label{fig:field} Results obtained from field trials conducted in one of the Northeast US regions. Due to proprietary restrictions, we cannot disclose the actual numbers of the volume, throughput, and the number of handovers. Instead, we normalized the values in this figure and provided the percentage.
    }
    \vspace{-0.2in}
\end{figure*}

\textbf{Evaluation Results:}
In the following, we present the evaluation results obtained by running the competing approaches in simulated scenarios.

\underline{Average Downlink UE Throughput}:
As the main KPI of the network, we evaluate the average aggregated downlink UE throughput of the proposed CaRL against other frameworks.
Fig.~\ref{fig:avg_tput_somerville} and \ref{fig:avg_tput_ohio} show the results of this metric for the two clusters. In cluster 1, we can observe that the heuristic algorithm can slightly improve the downlink throughput, but in the cluster 2 scenario, the performance of the heuristic algorithm is not so good.
This is because the heuristic algorithm performs a step-by-step adjustment, which has a much slower response time to network changes. Since the RAN status changes rapidly, this method cannot match the change rate of the RAN, and hence it cannot perform well. The difference in the performances of the heuristic algorithm in these two scenarios is caused by the different data volumes transmitted during the simulation. The volume of the data from cluster 2 is much larger than the volume in cluster 1; therefore, performance drops significantly. As for the Q-table method, it can perform well in cluster 1 because it contains the data collected from cluster 1, indicating a few missing matches during the table lookup. The performance drop of the Q-table in the cluster 2 scenario is expected because it does not contain any data from cluster 2. Therefore the Q-table is not very helpful.
In Fig.~\ref{fig:avg_tput_ohio}, we apply knowledge transfer to train a new policy for the new data from cluster 2. The results show that CaRL achieves the best performance among all others because of offline training.

\underline{Handovers}:
Even though load balancing is achieved by HOs, we argue that HOs are beneficial to the RAN only under certain conditions. Most HOs will lead to a transient drop in the QoE because of the temporary disconnection. Therefore, we evaluate the model on the number of HOs and use it as one of the main metrics to measure the QoE. Fig.~\ref{fig:total_ho_somerville} and \ref{fig:total_ho_ohio} present the total number of HOs during the simulation for the two scenarios. It can be observed that CaRL yields more HOs than other frameworks. However, we argue that, since the throughput of CaRL is improving, as shown in Fig.~\ref{fig:avg_tput_somerville} and \ref{fig:avg_tput_ohio}, the increased number of HOs in CaRL might just be the HOs that are helping to achieve load balancing. Furthermore, it is worth pointing out that even though the total number of HOs increased, individual users still do not perceive many HOs. Fig.~\ref{fig:ho_somerville} and \ref{fig:ho_ohio} show the distribution of the number of HOs per each UE. We can observe that most UEs in CaRL only have no more than 5 HOs in two hours, which is acceptable compared to other frameworks.

\underline{Sector-Level Performance Inspection}:
Load balancing is achieved by slightly sacrificing the high-quality users so that the low-quality users can be compensated. To visualize this process, 
we plot the number of RRC-connected UEs and the PRB utilization (i.e., the percentage of PRBs used over the total number of PRBs in a cell) at 15-minute intervals for three cells that are co-located in one cell sector in Fig.~\ref{fig:cells_nue} and \ref{fig:cells_prb}.
Since we only consider intra-sector HOs, therefore, for any one of these three cells, HOs are only possible if the target cells are also among these cells. We can read from the figures that CaRL can offload traffic from overloaded cells to less loaded cells. To illustrate, Cell 7C and Cell 9C are the overloaded cells in Fig.~\ref{fig:cells_nue_somerville} and \ref{fig:cells_nue_ohio}, respectively. The same can be observed from Fig.~\ref{fig:cells_prb_ohio}. CaRL tries to move traffic away from these two overloaded cells and use the other two cells as targets. The number of connected UEs between the results of CaRL and baseline is quite different, which further proves our claim that the increased number of HOs in Fig.~\ref{fig:ho_somerville} and \ref{fig:ho_ohio} are meant for improving the throughput. 
\underline{Ablation Study}:
To show the advantage of state space factorization and policy decomposition, we compare CaRL with the DDPG algorithm because if we set the number of sub-spaces to one, CaRL becomes DDPG. In Fig.~\ref{fig:sim_somerville} and \ref{fig:sim_ohio}, it can be observed that CaRL can outperform DDPG in both scenarios in terms of average aggregated downlink throughput. Although CaRL induces more HOs, these HOs are evenly distributed and are acceptable because most UEs have less than 5 HOs. Furthermore, in Fig.~\ref{fig:cells_nue} and \ref{fig:cells_prb}, we can also observe that CaRL can do a better job at load balancing. This advantage is more obvious in the cluster 2 scenario because the traffic patterns in this area are more complex (more UEs and heavier traffic).

\underline{Feature Importance revealed by CaRL}:
After jointly training the factorizer and the policies with policy gradient update, we now have a factorizer that can efficiently split the state space. To examine how well the factorizer's performance is, we plotted the importance the factorizer assigned to each feature in the state space after training. To do this, we pass the entire training dataset to the factorizer and generate a set of labels for each data sample. The dataset and the labels will then be used to train a simple decision tree, which will provide us with feature importance based on the data and labels. Fig.~\ref{fig:feature_imp} shows the feature importance. One may wonder why features like cell frequency and time of day are assigned much greater importance than the cell load.
The reason is that traffic patterns typically vary with factors like time and location; cell loads are just the results of the varied traffic patterns.


\vspace{-0.1in}
\subsection{Field Trial Evaluation}\label{sect:eval:field}

\textbf{Field Trial Setup:} To further evaluate the performance of CaRL in a real-world scenario, we applied CaRL to the recent field trial in the real RAN with a tier-1 RAN operator to see its performance in improving the average UE throughput. The field trials were carried out in one of the US Northeast regions, consisting of two areas, one near highways and the other near residential areas. Each testing area includes 3 LTE BSs with 12 cells, whose sector has four different bands. We selected two weeks (each from Tuesday to Saturday). One week is to collect the baseline (that is, not applying CaRL), and then, for the second week, we applied CaRL to see the improvement for the same areas by comparison. Besides, we found that the two areas had different traffic patterns, so we built two separate RL sub-policies, i.e., set $M=2$, and let the factorizer select the right one for a given input state. We reused an existing sub-policy for the high-way area using knowledge transferring.

\textbf{Results:} The results of the field trial are shown in Fig.~\ref{fig:field}. In Fig.~\ref{fig:field_vol}, the total traffic volume and its pattern for the field trial week are very similar to the baseline week, so we can consider that the evaluation of CaRL with the different week from the baseline week for the same areas is fair enough to see how CaRL deals with the similar traffic volume with the volume observed in the baseline week. Fig.~\ref{fig:field_thrpt} shows that CaRL has consistently improved throughput on all days. The overall average improvement is around 18.5\%. This figure also shows that improvement has been achieved during relatively busy hours (from 7 AM to 9 PM on each day). This is because cells during busy hours are under high load conditions and many controls have been triggered for load-balancing to improve throughput. 
Meanwhile, Fig.~\ref{fig:field_ho} shows that CaRL triggers more handovers than the baseline. It increases by about 57\%. 
It also shows that most such handovers have been triggered during busy hours because cells are easily overloaded, and CaRL has attempted load-balancing. We have carefully monitored the number of HOs to see if too many HOs incur any negative impact on users’ QoE with service retainability and accessibility. It turns out that their drop rate is just around 0.02\%, and almost all handovers have been successfully completed without failure even during peak times, so we can conclude that those extra handovers are acceptable.

\section{Conclusion} \label{sect:conclusion}

To reconcile the complexity of the traffic steering problem in the multi-cell setting, we introduced CaRL, a novel framework for traffic steering in Open-RAN networks. Our proposed state space factorization and policy decomposition approach efficiently handles large state spaces by classifying the incoming states into sub-spaces. We evaluated CaRL in a digital twin and demonstrated that it outperforms competing approaches by improving network throughput.
Furthermore, we conducted a field trial to show that the proposed CaRL can improve overall downlink throughput in a real-world setting.




\bibliographystyle{ieeetr}
\bibliography{csun}

\end{document}